# Calculation-Efficient Algorithms for Multiuser Detection in Short Code CDMA (Wideband CDMA Time Division Duplex) Systems.

Parthapratim De

Institute for Infocomm Research, Singapore.

e-mail:  pd4267@yahoo.com

**Abstract**

This paper derives and analyzes a novel block Fast Fourier Transform (FFT) based joint detection algorithm. The paper compares the performance and complexity of the novel block-FFT based joint detector to that of the Cholesky based joint detector and single user detection algorithms. The novel algorithm can operate at chip rate sampling, as well as higher sampling rates, unlike existing algorithms. For the performance/complexity analysis, the time division duplex (TDD) mode of a wideband code division multiplex access (WCDMA) is considered. The results indicate that the performance of the fast FFT based joint detector is comparable to that of the Cholesky based joint detector, and much superior to that of single user detection algorithms. On the other hand, the complexity of the fast FFT based joint detector is significantly lower than that of the Cholesky based joint detector and less than that of the single user detection algorithms. For the Cholesky based joint detector, the approximate Cholesky decomposition is applied. Moreover, the novel method can also be applied to any generic multiple-input-multiple-output (MIMO) system.

**I. Introduction**

In some communication systems, such as frequency division duplex CDMA (FDD/CDMA) and time division duplex CDMA (TDD/CDMA), multiple communications are sent over the same frequency spectrum. These communications are differentiated by their channelization (spreading) codes. TDD/CDMA communication systems use repeating frames divided into timeslots for communication. A single communication sent in such a system will have one or multiple associated codes and timeslots assigned to it. But for WCDMA systems [1], conventional RAKE receivers suffer from severe degradation in frequency selective fading channels because of significant multi-access interference and inter-symbol interference. For such systems, joint detector based multi-user detector algorithm [2], [3], [4] have attracted attention. Joint detector algorithms are characterized by good performance with high complexity. However, for short-code TDD/CDMA systems, codes have a length equal to the symbol period and lends themselves to the development of joint detectors. However spreading codes used in FDD/CDMA systems have a period much longer than the symbol period (called long spreading or scrambling codes), for which it is difficult to design multiuser detectors, as stated in [5].

Another approach to removing multi-access and inter symbol interference is single user detection [6], [7], [8]. This is an approach based on channel equalization and is applicable to the downlink of a CDMA system, without any transmit diversity. In single user detection, the received signal is passed through an equalization stage followed by de-spreading for recovering the data of a single mobile. The equalization stage can be implemented using an approximate Cholesky decomposition [3].

An earlier efficient FFT based implementation of the joint detector was proposed in [4]. However the approach in [4] is for chip rate sampling and cannot be extended easily to oversampling or multiple chip rate sampling, as explained later. Practical receivers typically operate at twice the chip rate or higher rates. This paper provides a novel fast joint detector algorithm that is applicable at any sampling rate. The fast implementation is achieved through a *block* circulant approximation of the correlation matrix.

This paper also shows that single user detection based algorithms exhibit some degradation, as compared to the joint detector algorithm of [2], [3] at the downlink of a TDD-CDMA system. One motivation to developing the fast joint detector algorithm in this paper was to develop an algorithm, low in complexity compared to the joint detector in [2], but superior in performance to that of the single user detection algorithms in [6], [7] at the downlink of a TD-CDMA system. Simulations are presented to illustrate the relative performance of the new algorithm for different multipath channels specified in [1]. The performance is analytically explained for different channel conditions.

Single user detection algorithms are also used in the case of multi-code transmission systems, where the data of a particular user may have been sent using multiple codes. WCDMA TDD mode has the option to support higher data rates, like the two Mbps data service for which twelve codes of spreading factor sixteen each are allocated to a single user. In this paper, we extend the new joint detector algorithm to the multi-code scenario. In particular, we study the performance of the new fast joint detector algorithm with that of the single user detector and other joint detector algorithms for this application. Computational complexities of the different algorithms are analyzed and it is seen that the novel algorithm in this paper provides far superior performance, at less computational complexity, compared to single user detection algorithms.

This paper also provides an analysis of the performance degradation of the single user detector algorithms in different multipath channels, and compares it to that of the fast joint detector algorithm in this paper. The paper also provides results at higher than chip rate sampling and illustrates the advantages of the new algorithm, as compared to the fast joint detector in [4]. The paper also considers some issues in the design of the fast joint detector algorithm, like the implementation of the inverse required in the method, along with an efficient prime factor algorithm (**PFA**) block-FFT implementation, which also improves the performance of the novel fast algorithm (called

*extended* fast joint detector). Some associated work is given in

 [9]-[17]. The novel method in this paper can also be applied to any generic multiple-input-multiple-output (MIMO) system like [18, [19].

The paper is organized as follows. Section II gives the signal model, while the algorithm derivation is given in Section III. Section IV provides details regarding implementation of the novel detector, as well as comparison of computational complexities among the detectors, along with comparison with single-user detectors. Simulation results are in Section V, while conclusions are included in Section VI.

**II. Signal Model**

 A typical Universal Terrestrial Radio Access (UTRA) WCDMA TDD communication burst is shown in Figure 1, which is a WCDMA system, in which the uplink and downlink transmissions are confined to different time-slots. Within each time-slot, multiple signals are multiplexed using CDMA. A typical communication burst has a midamble, a guard period and two data fields $D_1$ and $D_2$. The midamble separates the two data fields and contains a training sequence. The guard period separates the communication bursts to allow for the difference in arrival times of bursts transmitted from different transmitters. The two data fields contain the communication burst's data.

The fast joint detector in this paper is developed for this communication system. The receiver receives a combination of K bursts arriving simultaneously, which may be for K different mobiles (users). If certain mobiles are using multiple codes in a particular time slot, the K bursts may be for less than K users. Each data field of a timeslot has a predefined number of transmitted symbols, such as $N_S$. For the kth user, each of its $N_S$ symbols is spread by its code $\mathbf{c}^{(k)}$ (with spreading factor SF); accordingly, each data field has $N_S$ x SF chips. After passing through a channel having an impulse response of W chips, each received data field has a length of (SF x $N_s$ + W -1) chips, which is denoted by $N_c$ chips. A typical value of W is 57 chips. Each $k^{th}$ burst is received at the receiver and can be written as

$$\mathbf{r}^{(k)} = \mathbf{A}^{(k)}\mathbf{d}^{(k)}, k = 1,..,K . \qquad (1)$$

$\mathbf{r}^{(k)}$ is the received contribution of the $k^{th}$ burst; $\mathbf{A}^{(k)}$ is the combined channel response, being an $N_c$ x $N_s$ matrix. Each $j^{th}$ column in $A^{(k)}$ is a zero-padded and shifted version of the symbol response of the $j^{th}$ element of $\mathbf{d}^{(k)}$. The combined channel response is the convolution of the estimated multipath response $\mathbf{h}^{(k)}$ and spreading code $c^{(k)}$ for the burst. $\mathbf{d}^{(k)}$ is the unknown data symbols transmitted in the burst. The estimated multipath response for each $k^{th}$ burst, $\mathbf{h}^{(k)}$, has a length W. In the downlink, without any transmit diversity, all the bursts pass through the same channel $\mathbf{h}^{(k)}$ to a particular user. In uplink, the multipath responses $\mathbf{h}^{(k)}$ (from the different users to the base-station) are different. If transmit diversity is employed in downlink, then also $\mathbf{h}^{(k)}$ is different for each k. The overall received vector from all K bursts sent over the wireless channel is

$$\mathbf{r} = \sum_{k=1}^{K}\mathbf{r}^{(k)} + \mathbf{n} . \qquad (2)$$

### III Algorithm Development

By combining the $\mathbf{A}^{(k)}$'s for all the data bursts into matrix $\mathbf{A}$ and the unknown data for each user $\mathbf{d}^{(k)}$ into the vector $\mathbf{d}$, we have

$$\mathbf{r} = \mathbf{A}\mathbf{d} + \mathbf{n} . \qquad (3)$$

where $\mathbf{A}$ is of size ($N_c$) by ($N_s$.K) and $\mathbf{d}$ is of size ($N_s$.K) by 1. The above model includes both the multi-access interference and the intersymbol interference in the received signal. In equation (3), $\mathbf{r}$ is the chip rate sampled received vector of length $N_c$ chips and $\mathbf{n}$ is the zero-mean noise vector. The maximum number of users is K, with $b_j^{(k)}(i)$ being the convolution of the channel response $\mathbf{h}$ and the spreading code $\mathbf{c}^{(k)}$, for the kth user at the jth chip interval of the ith symbol interval. Define the (SF by K) block B(i) as

$$\mathbf{B}(i) = \begin{bmatrix} b_1^{(1)}(i) & \cdots & b_1^{(K)}(i) \\ \vdots & & \vdots \\ b_{SF}^{(1)}(i) & \cdots & b_{SF}^{(K)}(i) \end{bmatrix}. \qquad (4)$$

Now we can write the transfer matrix A, in (3), in terms of $\mathbf{B}(i)$.

$$\mathbf{A} = \begin{bmatrix} \mathbf{B}(0) & 0 & 0 & 0 & \cdots & 0 \\ \mathbf{B}(1) & \mathbf{B}(0) & 0 & 0 & \cdots & 0 \\ \mathbf{B}(2) & \mathbf{B}(1) & \mathbf{B}(0) & 0 & \cdots & 0 \\ \vdots & \vdots & \vdots & \vdots & \vdots & \vdots \\ 0 & 0 & \cdots & \mathbf{B}(2) & \mathbf{B}(1) & \mathbf{B}(0) \\ \vdots & \vdots & & \vdots & \vdots & \vdots \end{bmatrix}. \qquad (5)$$

Given that the data of each of the K users are uncorrelated (with zero mean and unity variance) and the data of any user is uncorrelated with the data of each of the other users, the correlation matrix of the received signal is given by $\mathbf{R} = \mathbf{A}^H\mathbf{A} + \sigma^2\mathbf{I}$. Then the minimum mean squared error (MMSE) solution to the joint detection problem is given by

$$\mathbf{R}\hat{\mathbf{d}} = (\mathbf{A}^H\mathbf{A} + \sigma^2\mathbf{I})\hat{\mathbf{d}} = \mathbf{A}^H\mathbf{r}. \qquad (6)$$

The matrix $\mathbf{R}$ is block Toeplitz as well as banded and algorithms for banded block Toeplitz system can be applied to it (e.g. approximate Cholesky joint detection [3]). For illustration purposes, a simplified example of $\mathbf{R}$ with $N_s = 10$ and $W = 2$ is given by

$$\mathbf{R} = \begin{bmatrix} R_0 & R_1^H & R_2^H & 0 & 0 & 0 & 0 & 0 & 0 & 0 \\ R_1 & R_0 & R_1^H & R_2^H & 0 & 0 & 0 & 0 & 0 & 0 \\ \cdots & \cdots & \cdots & \cdots & \cdots & \cdots & \cdots & \cdots & \cdots & \cdots \\ R_2 & R_1 & R_0 & R_1^H & R_2^H & 0 & 0 & 0 & 0 & 0 \\ 0 & R_2 & R_1 & R_0 & R_1^H & R_2^H & 0 & 0 & 0 & 0 \\ 0 & 0 & R_2 & R_1 & R_0 & R_1^H & R_2^H & 0 & 0 & 0 \\ 0 & 0 & 0 & R_2 & R_1 & R_0 & R_1^H & R_2^H & 0 & 0 \\ 0 & 0 & 0 & 0 & R_2 & R_1 & R_0 & R_1^H & R_2^H & 0 \\ 0 & 0 & 0 & 0 & 0 & R_2 & R_1 & R_0 & R_1^H & R_2^H \\ \cdots & \cdots & \cdots & \cdots & \cdots & \cdots & \cdots & \cdots & \cdots & \cdots \\ 0 & 0 & 0 & 0 & 0 & 0 & R_2 & R_1 & R_0 & R_1^H \\ 0 & 0 & 0 & 0 & 0 & 0 & 0 & R_2 & R_1 & R_0 \end{bmatrix}. \qquad (7)$$

This example is easily extendable for any $N_s$ and W. The matrix $\mathbf{R}$ is of size $(KN_s)$ by $(KN_s)$, each entry $\mathbf{R}_i$ being a K by K block. The matrix **R** has an additional structure- the banded and Toeplitz structure of R makes the submatrix, within the dotted lines of **R**, block-circulant. The portion of **R**, which is not block-circulant, depends on the maximum multipath delay spread W (in chips).

Practical receivers for TDD system operate at twice the chip rate or a multiple of the chip rate to provide robustness to timing errors. For multiple chip rate sampling, the received signal model is

$$\begin{bmatrix} \mathbf{r}_1 \\ \mathbf{r}_2 \\ \vdots \\ \mathbf{r}_N \end{bmatrix} = \begin{bmatrix} \mathbf{A}_1 \\ \mathbf{A}_2 \\ \vdots \\ \mathbf{A}_N \end{bmatrix} \mathbf{d} + \mathbf{n} \qquad (8)$$

where sampling is at N times the chip rate. The sequence $\mathbf{r}_n$ corresponds to the nth chip rate sub-vector, out of N such sequences corresponding to oversampling by a factor of N. In this case, the correlation matrix **R** can be written as

$$\mathbf{R} = \sum_{i=1}^{N} \mathbf{A}_i^H \mathbf{A}_i + \sigma^2 \mathbf{I}. \qquad (9)$$

The structure of the **R** matrix for multiple chip rate sampling is still the same as in equation (7). The number of row-blocks of **R** that is outside the block circulant structure is equal to 2L, where L is the multipath delay spread (in symbols) where $L = (SF + W - 1)/SF$. The algorithm derived in this paper is based on block-circulant extension of the matrix **R**. Since a block circulant approximation of the correlation matrix R, in equation (9), is used, the fast joint detector can be easily developed for higher rate sampling. This is in contrast with the approach in [4], where a circulant approximation to the data matrix (**not** correlation matrix) is considered and requires addition of block columns to make it block square. In the case of oversampling given by the model in equation (8), this will be cumbersome. The approach in this paper has no such problems and can be extended to the oversampling case in a straight forward manner.

The derivation of a block FFT based joint detector is outlined below. This will enable appreciation of the final form of the algorithm. The block circulant extension of the correlation matrix **R** is denoted by $\mathbf{R}_c$.

$$\mathbf{R}_c = \begin{bmatrix} R_0 & R_1^H & R_2^H & 0 & 0 & 0 & 0 & 0 & R_2 & R_1 \\ R_1 & R_0 & R_1^H & R_2^H & 0 & 0 & 0 & 0 & 0 & R_2 \\ \cdots & \cdots & \cdots & \cdots & \cdots & \cdots & \cdots & \cdots & \cdots & \cdots \\ R_2 & R_1 & R_0 & R_1^H & R_2^H & 0 & 0 & 0 & 0 & 0 \\ 0 & R_2 & R_1 & R_0 & R_1^H & R_2^H & 0 & 0 & 0 & 0 \\ 0 & 0 & R_2 & R_1 & R_0 & R_1^H & R_2^H & 0 & 0 & 0 \\ 0 & 0 & 0 & R_2 & R_1 & R_0 & R_1^H & R_2^H & 0 & 0 \\ 0 & 0 & 0 & 0 & R_2 & R_1 & R_0 & R_1^H & R_2^H & 0 \\ 0 & 0 & 0 & 0 & 0 & R_2 & R_1 & R_0 & R_1^H & R_2^H \\ \cdots & \cdots & \cdots & \cdots & \cdots & \cdots & \cdots & \cdots & \cdots & \cdots \\ R_2^H & 0 & 0 & 0 & 0 & 0 & R_2 & R_1 & R_0 & R_1^H \\ R_1^H & R_2^H & 0 & 0 & 0 & 0 & 0 & R_2 & R_1 & R_0 \end{bmatrix} \quad (10)$$

Let a matrix $\mathbf{D}$ be determined such that $\mathbf{R}_c = \mathbf{D}\Lambda\mathbf{D}^H$. It is shown below that

$$\mathbf{D} = \begin{bmatrix} I & e^{\frac{j2\pi}{N_s}}I_K & e^{\frac{j4\pi}{N_s}}I_K & \cdots \cdots & e^{\frac{j2(N_s-1)\pi}{N_s}}I_K \\ I & e^{\frac{j4\pi}{N_s}}I_K & e^{\frac{j8\pi}{N_s}}I_K & \cdots \cdots & e^{\frac{j4(N_s-1)\pi}{N_s}}I_K \\ \vdots & \vdots & \vdots & & \vdots \\ I & e^{\frac{j2(N_s-1)\pi}{N_s}}I_K & e^{\frac{j4(N_s-1)\pi}{N_s}}I_K & \cdots \cdots & e^{\frac{j2(N_s-1)^2\pi}{N_s}}I_K \\ I & e^{\frac{j2N_s\pi}{N_s}}I_K & e^{\frac{j4N_s\pi}{N_s}}I_K & \cdots \cdots & e^{\frac{j2(N_s-1)N_s\pi}{N_s}}I_K \end{bmatrix} \quad (11)$$

where $I_K$ is a K by K identity matrix. $\mathbf{D}$ is a ``block Discrete Fourier Transform (DFT)''– like matrix, i.e. each of its entry is a (K by K) block. Furthermore, it can be shown that $\mathbf{D}^H\mathbf{D} = N_s I_{KN_s}$, where $I_{KN_s}$ is the $KN_s$ by $KN_s$ identity matrix. Let the block-diagonal matrix $\Lambda$, of size ($KN_s$ by $KN_s$), be

$$\Lambda \stackrel{\Delta}{=} \begin{bmatrix} \Lambda^{(1)} & & & & \\ & \Lambda^{(2)} & & & \\ & & \Lambda^{(3)} & & \\ & & & \ddots & \\ & & & & \Lambda^{(N_s)} \end{bmatrix}, \quad (12)$$

with each entry

$$\Lambda^{(i)} = \begin{bmatrix} \lambda_{11}^{(i)} & \cdots & \lambda_{1K}^{(i)} \\ \vdots & & \vdots \\ \lambda_{K1}^{(i)} & \cdots & \lambda_{KK}^{(i)} \end{bmatrix} \qquad (13)$$

of size (K by K). Next, the block-circulant matrix $\mathbf{R}_c$ is multiplied by the matrix $\mathbf{D}$ to form the matrix $\mathbf{R}_c\mathbf{D}$, with (K by K) blocks as entries; this is shown (below, in page 10) in equation (14). Similarly, the $\mathbf{D}$ matrix is multiplied by the $\Lambda$ matrix in (15). The system of equations obtained by equating each block-row of $\mathbf{R}_c\mathbf{D}$ with the same row of $\mathbf{D}\Lambda$ is consistent.

$$\mathbf{D}\Lambda = \begin{bmatrix} \Lambda^{(1)} & \Lambda^{(2)} e^{j\frac{2\pi}{N_s}} I_K & \Lambda^{(3)} e^{j\frac{4\pi}{N_s}} I_K & \cdots & \cdots & \Lambda^{(N_s)} e^{j\frac{2(N_s-1)\pi}{N_s}} I_K \\ \Lambda^{(1)} & \Lambda^{(2)} e^{j\frac{4\pi}{N_s}} I_K & \Lambda^{(3)} e^{j\frac{8\pi}{N_s}} I_K & \cdots & \cdots & \Lambda^{(N_s)} e^{j\frac{4(N_s-1)\pi}{N_s}} I_K \\ \vdots & \vdots & \vdots & & & \vdots \\ \Lambda^{(1)} & \Lambda^{(2)} e^{j\frac{2(N_s-1)\pi}{N_s}} I_K & \Lambda^{(3)} e^{j\frac{4(N_s-1)\pi}{N_s}} I_K & \cdots & \cdots & \Lambda^{(N_s)} e^{j\frac{2(N_s-1)^2\pi}{N_s}} I_K \\ \Lambda^{(1)} & \Lambda^{(2)} e^{j\frac{2N_s\pi}{N_s}} I_K & \Lambda^{(3)} e^{j\frac{4N_s\pi}{N_s}} I_K & \cdots & \cdots & \Lambda^{(N_s)} e^{j\frac{2(N_s-1)N_s\pi}{N_s}} I_K \end{bmatrix} . \quad (15)$$

Accordingly, the same set of equations are obtained by equating any block-row of $\mathbf{R}_c\mathbf{D}$ with the same block-row of $\mathbf{D}\Lambda$. To illustrate for equation (14), the first block-row of $\mathbf{R}_c\mathbf{D}$ is given by

$$\begin{aligned} \big[ & (\mathbf{R}_0 + \mathbf{R}_1 + \mathbf{R}_2 + \mathbf{R}_1^H + \mathbf{R}_2^H), \\ & (\mathbf{R}_0 e^{j\frac{2\Pi}{N_s}} + \mathbf{R}_1^H e^{j\frac{4\Pi}{N_s}} + \mathbf{R}_2^H e^{j\frac{6\Pi}{N_s}} + \mathbf{R}_2 e^{j\frac{18\Pi}{N_s}} + \mathbf{R}_1 e^{j\frac{20\Pi}{N_s}}), \\ & (\mathbf{R}_0 e^{j\frac{4\Pi}{N_s}} + \mathbf{R}_1^H e^{j\frac{8\Pi}{N_s}} + \mathbf{R}_2^H e^{j\frac{12\Pi}{N_s}} + \mathbf{R}_2 e^{j\frac{36\Pi}{N_s}} + \mathbf{R}_1 e^{j\frac{40\Pi}{N_s}}), \ldots \\ & (\mathbf{R}_0 e^{j\frac{2(N_s-1)\pi}{N_s}} + \mathbf{R}_1^H e^{j\frac{4(N_s-1)\pi}{N_s}} + \mathbf{R}_2^H e^{j\frac{6(N_s-1)\pi}{N_s}} + \mathbf{R}_2 e^{j\frac{2(N_s-1)^2\pi}{N_s}} \\ & + \mathbf{R}_1 e^{j\frac{2(N_s-1)N_s\pi}{N_s}}). \big] \end{aligned}$$

(16)

Similarly, the first block-row of $\mathbf{D}\Lambda$ is given by

$$\left[ \Lambda^{(1)}, \Lambda^{(2)} e^{\frac{j2\pi}{N_s}} I_K, \Lambda^{(3)} e^{\frac{j4\pi}{N_s}} I_K, \ldots, \Lambda^{(N_s)} e^{\frac{j2(N_s-1)\pi}{N_s}} I_K \right] . \quad (17)$$

Equating (16) and (17), we have the solution of $\Lambda^{(k)}$'s,

$$\Lambda^{(1)} = (R_0 + R_1 + R_2 + R_1^H + R_2^H), \qquad (18)$$

$$\Lambda^{(2)} e^{\frac{j2\pi}{N_s}} I_K = (R_0 e^{j\frac{2\Pi}{N_s}} + R_1^H e^{j\frac{4\Pi}{N_s}} + R_2^H e^{j\frac{6\Pi}{N_s}} + R_2 e^{j\frac{18\Pi}{N_s}} + R_1 e^{j\frac{20\Pi}{N_s}})$$

$$\Lambda^{(2)} = (R_0 + R_1^H e^{j\frac{2\Pi}{N_s}} + R_2^H e^{j\frac{4\Pi}{N_s}} + R_2 e^{-j\frac{4\Pi}{N_s}} + R_1 e^{-j\frac{2\Pi}{N_s}}).$$

(19)

Similarly, $\Lambda^{(N_s-1)}$ can be calculated as

$$\Lambda^{(N_s-1)} = (R_0 + R_1^H e^{j\frac{2(N_s-2)\Pi}{N_s}} + R_2^H e^{j\frac{4(N_s-2)\Pi}{N_s}} + R_2 e^{-j\frac{4(N_s-2)\Pi}{N_s}}$$

$$+ R_1 e^{-j\frac{2(N_s-2)\Pi}{N_s}})$$

(20)

and

$$\Lambda^{(N_s)} = (R_0 + R_1^H e^{j\frac{2(N_s-1)\Pi}{N_s}} + R_2^H e^{j\frac{4(N_s-1)\Pi}{N_s}} + R_2 e^{-j\frac{4(N_s-1)\Pi}{N_s}}$$

$$+ R_1 e^{-j\frac{2(N_s-1)\Pi}{N_s}}).$$

(21)

Although equations (16) to (21) are illustrated using the first block-row of $R_cD$ and $D\Lambda$, the same expressions will be obtained using any block-row of $R_cD$ and $D\Lambda$. Same results are obtained by using center block-row ($(N_s/2)$th block-row) in equation (14).

Equations (18) to (21) compute the block FFTs of K by K blocks. Since these blocks are multiplied by scalar exponentials, this process is referred to as ``block''-FFTs. $\Lambda^{(k)}$ can be obtained by computing the block-FFT of $\{\mathbf{0}, \mathbf{0}, \ldots, \mathbf{R}_2, \mathbf{R}_1, \mathbf{R}_0, \mathbf{R}_1^H, \mathbf{R}_2^H, \ldots, \mathbf{0}, \mathbf{0}\}$. As shown in the above equations (18) to (21), all the $\Lambda^{(k)}$'s can be calculated using one block-row of $\mathbf{R}$; thus $\mathbf{R}_c$ does not need to be explicitly computed. Any block-row of $\mathbf{R}$, which is at least L (L is the delay-spread of the channel in symbols) block-rows from both the top and the bottom of the $\mathbf{R}$ matrix, can be used to calculate $\Lambda^{(k)}$'s, since these block-rows have the full set of $\mathbf{R}_i$'s. In general form, we have

$$\Lambda^{(k)} = (\mathbf{R}_L e^{\frac{-j2L(k-1)\pi}{N_S}} + \ldots + \mathbf{R}_1 e^{\frac{-j2(k-1)\pi}{N_S}} + \mathbf{R}_0 + \mathbf{R}_1^H e^{\frac{j2(k-1)\pi}{N_S}} + \ldots$$
$$+ \mathbf{R}_L^H e^{\frac{j2L(k-1)\pi}{N_S}}), \quad k = 0,1,\ldots,N_S. \quad (22)$$

Thus $\Lambda^{(k)}$ can be calculated as the FFT of a *two-sided* sequence of (K by K) blocks $\{\mathbf{R}_0, \mathbf{R}_1, \mathbf{R}_2, \ldots, \mathbf{R}_L\}$. The above equation thus requires calculation of $K^2$ FFTs, each of length $N_s$. Using the $\Lambda^{(k)}$ and $\mathbf{D}$ matrices, the block-circulant matrix $\mathbf{R}_c$ can be written as

$$\mathbf{R}_c \mathbf{D} = \mathbf{D}\Lambda, \quad (23)$$

and

$$\mathbf{R}_c = \frac{1}{N_S} \mathbf{D}\Lambda\mathbf{D}^H. \quad (24)$$

$\mathbf{D}$ and $\Lambda$ are each of size $KN_s$ by $KN_s$. Since $\mathbf{D}^H\mathbf{D} = N_s\mathbf{I}_{KN_S}$, $\mathbf{D}^{-1} = \frac{1}{N_S}\mathbf{D}^H$. Thus we have

$$\mathbf{R}_c^{-1} = N_S[(\mathbf{D}^H)^{-1}\Lambda^{-1}\mathbf{D}^{-1}] = N_S[(\mathbf{D}/N_S)\Lambda^{-1}(\mathbf{D}^H/N_S)]. \quad (25)$$

The detected data vector $\hat{\mathbf{d}}$ (of size $(KN_s)$ by 1) can be estimated using the MMSE criterion as

$$\hat{\mathbf{d}} = \mathbf{R}_c^{-1}(\mathbf{A}^H\mathbf{r}) = (\mathbf{D}/N_S)\Lambda^{-1}\mathbf{D}^H(\mathbf{A}^H\mathbf{r}) \quad (26)$$

which can be written as

$$\mathbf{D}^H \hat{\mathbf{d}} = \Lambda^{-1}[\mathbf{D}^H(\mathbf{A}^H \mathbf{r})]. \qquad (27)$$

The matrix $\Lambda$ (of size ($KN_s$ by $KN_s$)) is a block-diagonal matrix, with each block of being size of (K by K).

$$\Lambda = \begin{bmatrix} \Lambda^{(1)} & & & & \\ & \Lambda^{(2)} & & & \\ & & \Lambda^{(3)} & & \\ & & & \ddots & \\ & & & & \Lambda^{(N_s)} \end{bmatrix}.$$

Thus

$$\Lambda^{-1} = \begin{bmatrix} \Lambda^{(1)} & & & & \\ & \Lambda^{(2)} & & & \\ & & \Lambda^{(3)} & & \\ & & & \ddots & \\ & & & & \Lambda^{(N_s)} \end{bmatrix}^{-1}$$

$$= \begin{bmatrix} [\Lambda^{(1)}]^{-1} & & & & \\ & [\Lambda^{(2)}]^{-1} & & & \\ & & [\Lambda^{(3)}]^{-1} & & \\ & & & \ddots & \\ & & & & [\Lambda^{(N_s)}]^{-1} \end{bmatrix}. \qquad (28)$$

The inversion requires $N_s$ inversions of (K by K) matrices $\Lambda^{(k)}$. As a result, the data estimation equation (27) can be re-written as

$$[F(\hat{\mathbf{d}})]_k = [\Lambda^{(k)}]^{-1}[F(\mathbf{A}^H \mathbf{r})]_k \qquad (29)$$

for each frequency point k. F(.) refers to the FFT operation. Equation (29) is applicable to both receivers that sample the received signal at the chip rate and those that oversample the received signal at a multiple of the chip rate, such as twice the chip rate. For multiple chip rate receivers, the matrix **R**, corresponding both to multiple chip rate sampling or just chip rate signaling, is of the form as equation (7), being approximately block-circulant.

Further reductions in complexity in computing the matched filter ($\mathbf{A}^H \mathbf{r}$) can be achieved by exploiting the special structure of the overall channel response matrix **A**, which is again exploited by the FFT method. A glance at the matrices **A** and B(i) in equations (5) and (4) shows that the matrix **A** (of size $N_S SF \times N_S K$), is a non-square matrix but a portion of **A** is also block-circulant. L block-rows at the top and L block-rows at the bottom of the matrix A prevent it from being circulant. A block-circulant extension of the matrix A is denoted by $\mathbf{A}_c$. This matrix can be written as

$$\mathbf{A}_c = \mathbf{D}_1 \mathbf{\Lambda}_1 \mathbf{D}_2^H, \qquad (30)$$

where $\mathbf{D}_1$ is a ($N_S SF \times N_S SF$) matrix, $\mathbf{D}_2$ is a ($N_S K \times N_S K$) matrix and $\mathbf{\Lambda}_1$ is a ($N_S SF \times N_S K$) block-diagonal matrix. The matrix $\mathbf{\Lambda}_1$ is of the same form as $\mathbf{\Lambda}$ with ($N_S$ by $N_S$) blocks, except that each block $\mathbf{\Lambda}_1^{(i)}$ is non-square, being of size SF by K, and

$$\mathbf{D}_1 = \begin{bmatrix} \mathbf{I} & e^{\frac{j2\pi}{N_S}} \mathbf{I}_{SF} & e^{\frac{j4\pi}{N_S}} \mathbf{I}_{SF} & \cdots & \cdots & e^{\frac{j2(N_S-1)\pi}{N_S}} \mathbf{I}_{SF} \\ \mathbf{I} & e^{\frac{j4\pi}{N_S}} \mathbf{I}_{SF} & e^{\frac{j8\pi}{N_S}} \mathbf{I}_{SF} & \cdots & \cdots & e^{\frac{j4(N_S-1)\pi}{N_S}} \mathbf{I}_{SF} \\ \vdots & \vdots & \vdots & & & \vdots \\ \mathbf{I} & e^{\frac{j2(N_S-1)\pi}{N_S}} \mathbf{I}_{SF} & e^{\frac{j4(N_S-1)\pi}{N_S}} \mathbf{I}_{SF} & \cdots & \cdots & e^{\frac{j2(N_S-1)^2\pi}{N_S}} \mathbf{I}_{SF} \\ \mathbf{I} & e^{\frac{j2N_S\pi}{N_S}} \mathbf{I}_{SF} & e^{\frac{j4N_S\pi}{N_S}} \mathbf{I}_{SF} & \cdots & \cdots & e^{\frac{j2(N_S-1)N_S\pi}{N_S}} \mathbf{I}_{SF} \end{bmatrix}.$$

(31)

$\mathbf{D}_2$ is of the same form as the matrix **D** in equation (11). Post-multiplying $\mathbf{A}_c$ by $\mathbf{D}_2$, it is seen (below in equation (32) in page 11), that blocks of size SF by K are formed. Similarly, in multiplying $\mathbf{D}_1$ and $\mathbf{\Lambda}_1$, products are formed of size $N_S SF$ by $N_S K$, each block of SF by K. Comparing any row of $\mathbf{A}_c \mathbf{D}_2$ with any row of $\mathbf{D}_1 \mathbf{\Lambda}_1$, (33) is obtained

$$\mathbf{\Lambda}_1^{(1)} = [\mathbf{B}(0) + \mathbf{B}(1) + \mathbf{B}(2)],$$
$$\mathbf{\Lambda}_1^{(2)} = [\mathbf{B}(0) + \mathbf{B}(1) e^{\frac{-j2\Pi}{N_S}} + \mathbf{B}(2) e^{\frac{-j4\Pi}{N_S}}],$$
$$\vdots$$
$$\mathbf{\Lambda}_1^{(N_S-1)} = [\mathbf{B}(0) + \mathbf{B}(1) e^{\frac{-j2(N_S-2)\Pi}{N_S}} + \mathbf{B}(2) e^{\frac{-j4(N_S-2)\Pi}{N_S}}],$$
$$\mathbf{\Lambda}_1^{(N_S)} = [\mathbf{B}(0) + \mathbf{B}(1) e^{\frac{-j2(N_S-1)\Pi}{N_S}} + \mathbf{B}(2) e^{\frac{-j4(N_S-1)\Pi}{N_S}}].$$

(33)

As a result, each $\Lambda_1^{(k)}$ can be determined as the *one-sided* FFT of (SF by K) blocks $\mathbf{B}(i)$'s. From equation (30) and using the fact that $\mathbf{D}_2^H \mathbf{D}_2 = N_s \mathbf{I}_{KN_s}$, we have

$$\mathbf{A}_c^H \mathbf{r} = \mathbf{D}_2 \Lambda_1^H (\mathbf{D}_1^H \mathbf{r}), \qquad (34)$$

$$\mathbf{D}_2^H (\mathbf{A}_c^H \mathbf{r}) = N_s [\Lambda_1]^H ((\mathbf{D}_1^H \mathbf{r})). \qquad (35)$$

Accordingly, $F(\mathbf{A}_c^H \mathbf{r})_k$ can be determined using FFTs per

$$[F(\mathbf{A}_c^H \mathbf{r})]_k = N_s [\Lambda_1^{(k)}]^H [(F(\mathbf{r})]_k. \qquad (36)$$

Similarly, since the matrix A is approximately block-circulant,

$\mathbf{R} = (\mathbf{A}^H \mathbf{A} + \sigma^2 \mathbf{I})$ can be implemented using FFTs using $\Lambda_1$.

The complexity in implementing the data estimation algorithm is considered next.

$$[F(\hat{\mathbf{d}})]_k = [\Lambda^{(k)}]^{-1} [F(\mathbf{A}^H \mathbf{r})]_k. \qquad (37)$$

The matrix $[\Lambda^{(k)}]$ is Hermitian but has no special structure. To reduce computational complexity of calculating of each $[\Lambda^{(k)}]^{-1}$, the solution of the above linear system can be performed using a LU (lower-upper) decomposition. Each $[\Lambda^{(k)}]$ is a (K by K) matrix whose LU decomposition is given by

$$\Lambda^{(k)} = \mathbf{LU} \qquad (38)$$

where **L** is a lower triangular matrix and **U** is an upper triangular matrix. Equation (37) can then be solved by using forward and backward substitution. The forward substitution uses the lower triangular matrix **L** as (starting from equation (37))

$$\begin{aligned}
&[\Lambda^{(k)}][F(\hat{\mathbf{d}})]_k = [F(\mathbf{A}^H \mathbf{r})]_k \Rightarrow \mathbf{LU}[F(\hat{\mathbf{d}})]_k = [F(\mathbf{A}^H \mathbf{r})]_k \\
&\Rightarrow \mathbf{Ly} = [F(\mathbf{A}^H \mathbf{r})]_k
\end{aligned} \qquad (39)$$

where $\mathbf{y} = \mathbf{U}[F(\hat{\mathbf{d}})]_k$. The backward substitution then uses the upper triangular matrix $\mathbf{U}$ to solve for $[F(\hat{\mathbf{d}})]_k$ by

$$\mathbf{U}[F(\hat{\mathbf{d}})]_k = \mathbf{y}. \qquad (40)$$

To improve the bit error rate (BER) for the data symbols at the ends of each data field, samples from the midamble portion and the guard period are used in the data estimation algorithm, as shown in Figure 1. To collect all the samples of the last symbols in the data fields, the samples $\mathbf{r}$ are extended by W-1 chips (length of channel impulse response) into the midamble and guard period. This permits utilization of all the multipath components of the last data symbols (of each data field) for data estimation purposes. The known midamble sequences are cancelled from the received samples (corresponding to the midamble), prior to data estimation. Similarly for the guard period for the second data field.

**IV. A Implementation Issues of the Fast Joint Detector**

One very important implementation issue is regarding the implementation of the block FFTs. In a TDD burst, the midamble has 512 chips whereas each of the two data fields has 976 chips, which at the spreading factor of 16, corresponds to 61 symbols. Implementation of equation (22) requires implementing a 61- point block FFT, where each block is of size (K by K). This would require implementation of a prime factor algorithm (**PFA**) for block-FFT implementation and would require more computations than the implementation of a $2^n$-point FFT. One solution to this is to increase the processing length to 64 symbols. The received signal vector $\mathbf{r}$, instead of being 61 symbols long for the data field of a timeslot, will be 64 symbols long by extending it into the midamble field and guard period. Simulations indicate the performance of this *extended* FFT joint detector improves, particularly for Case 2 channel. This can be explained from the structure of the matrix $\mathbf{R}$ in equation (7) (and from its block-circulant version in (10)), which illustrates the degradation due to the block-FFT approximation. For a given channel, the number of block-rows outside the block circulant portion of $\mathbf{R}$ is fixed at 2L. If the processing length is $N_s = 61$, then the size of the correlation matrix $\mathbf{R}$ is $(KN_s)$ by $(KN_s)$, out of which 2KL rows are outside the circulant matrix portion. When we increase the processing length to $(N_s+3)$, the size of $\mathbf{R}$ is $[K \cdot (N_s+3)]$ by $[K \cdot (N_s+3)]$. However, the number of block-rows outside the block circulant portion of $\mathbf{R}$ remains fixed at 2L. Thus the fraction of non-circulant rows in $\mathbf{R}$ (for $N_s = 61$) is $(2L/N_s)$. However, when we increase the processing to 64, (so as to perform $2^n$-point block-FFT instead of computationally cumbersome PFA version of FFT), the fraction of non-circulant rows in $\mathbf{R}$ is $(2L/(N_s + 3)) = (2L/(64))$, which is less than $(2L/N_s)$ (for the PFA version of FFT). Thus, the circulant approximation becomes more accurate when we increase the processing length to enable efficient implementation of the block FFTs. The increase of the processing length is shown in Figure 1.

## IV. B. Higher Data Rate Services

For higher data rate services, WCDMA TDD uses a multi-code, multi-slot strategy. For 2 Mbps data service, twelve codes of spreading factor 16 each and twelve out of the 15 timeslots in a radio frame are allocated to one user. Refer to equation (4) for the entries of matrix B which are used to form the system matrix A. If M codes are allocated to one user, then M columns of the matrix B are allocated to that user while the remaining (K-M) columns of the matrix B are for the remaining users. Simulation results are shown in this paper for the 2 Mbps data service.

## IV. C. Computational Complexity

In this section, we introduce some other data estimation algorithms for comparison purposes. The approximate Cholesky based joint detector [3] is commonly used in a TDD system-it performs very well with achievable complexity. An approximate Cholesky decomposition of the matrix R is extended to obtain the full Cholesky factor G such that $R = GG^H$. Let us denote this algorithm by JDChol. Single user detection algorithms developed in [6], [7] are designed for the downlink situation. In the downlink, without any transmit diversity, (for an oversampling factor of N), the received signal can be written as (starting from equation (8)),

$$\begin{bmatrix} \mathbf{r}_1 \\ \mathbf{r}_2 \\ \vdots \\ \mathbf{r}_N \end{bmatrix} = \begin{bmatrix} \mathbf{A}_1 \\ \mathbf{A}_2 \\ \vdots \\ \mathbf{A}_N \end{bmatrix} \mathbf{d} + \mathbf{n} = \begin{bmatrix} \mathbf{H}_1 \\ \mathbf{H}_2 \\ \vdots \\ \mathbf{H}_N \end{bmatrix} \begin{bmatrix} \mathbf{C}_1 & \mathbf{C}_2 & \ldots & \mathbf{C}_K \end{bmatrix} \mathbf{d} + \mathbf{n} .$$

(41)

The reason is as follows. $\mathbf{A}_i$ (for the ith oversampling branch) contains the convolution of the estimated response $\mathbf{h}_i^{(k)}$ and spreading code $\mathbf{c}^{(k)}$ the (for kth user) for the burst. $\mathbf{d}^{(k)}$ is the unknown data symbols transmitted in the burst. In the downlink, without any transmit diversity, all the bursts pass (from the base-station) through the same channel $\mathbf{h}^{(k)} = \mathbf{h}$ to a particular user. In equation (41), $\mathbf{C}_k$ is the associated code matrix for kth user, as in [5]. $\mathbf{H}_i$ is the associated channel matrix for the ith oversampling branch, and is the same for all the users. Let $\mathbf{s} = \begin{bmatrix} \mathbf{C}_1 & \mathbf{C}_2 & \ldots & \mathbf{C}_K \end{bmatrix} \mathbf{d}$. $\mathbf{s}$ can be estimated by the MMSE criterion,

$$\hat{\mathbf{s}} = (\mathbf{H}^H \mathbf{H} + \sigma^2 \mathbf{I})^{-1} (\mathbf{H}^H \mathbf{r}),  \qquad (42) \quad \text{where}$$

$$\mathbf{H} = [\mathbf{H}_1^H \mathbf{H}_2^H \cdots \mathbf{H}_N^H]^H . \qquad (43)$$

And then de-spreading $\hat{s}$ (by the code-matrix $[C_1 \ C_2 \ \ldots \ C_K]$) to obtain estimate of data symbols (of all K users) $\hat{d}$. For the FFT version of the single user detector (SDFFT), it is observed that

$$\tilde{R} = HH^H = \sum_{i=1}^{N} H_i H_i^H$$

is also approximately circulant (**not** block-circulant) and the portion, that is not circulant, is equal to 2W rows. A derivation, similar to above, gives us

$$F(\hat{s}) = \frac{F(H^H r)}{N_s F((\tilde{R})_i)} \qquad (44)$$

where $F((\tilde{R})_i))$ is the *scalar* Fourier transform of a suitable column of $(\tilde{R})_i$. Then $\hat{d}$ is obtained by de-spreading $\hat{s}$. This algorithm is denoted by SDFFT.

Another single-user detection algorithm is obtained by the application of approximate Cholesky based method in [3] to $(H^H H + \sigma^2 I)^{-1}$ in (chip-level) equalization stage (in equation (42)), followed by de-spreading. This algorithm is denoted by SDChol.

An analysis of the computational complexity of the joint detector and its comparison with other data detectors is now undertaken. The complexity of calculating $A$ is $K.SF.W$. The computational complexity of calculating $A^H A$ is

$$\frac{(K^2 + K)[2(SF + W - 1) - (n_{max} - 1)]}{2} \frac{n_{max}}{2}$$

$$-\frac{(K^2 - K)(SF + W - 1)}{2}$$

where $n_{max} = \min(N_s, (SF + W - 1)/SF + 1)$. Calculating $A^H r$ (a matrix vector calculation) requires $K N_s (SF + W - 1)$ calculations. Calculating the FFT of the jth block-column of R requires $K^2 (N_s \log_2 N_s)$. The inversion of each matrix $[\Lambda^{(k)}]$, without LU decomposition, requires $K^3$ calculations. For $N_s$ frequency points, the total number of calculations is $N_s K^3$. Calculating

$[F(\hat{d})]_k = [\Lambda^{(k)}]^{-1} [F(A^H r)]_k$ requires $K^2$ each for $N_s$ frequency points, resulting in $N_s K^2$ total number of calculations. The inverse FFT of $[F(\hat{d})]$ requires $K(N_s \log_2 N_s)$ computations.

To illustrate the complexity for fast joint (multiuser) detection, the number of million real operations per second (MROPS) for processing a TDD Burst Type I [1], with $N_c = 976$, $N_s = 61$. spreading factor $SF = 16$, number of codes (users) $K = 8$, is determined. The multipath delay spread W is 57 chips. The calculations of $\mathbf{A}$, $\mathbf{A}^H\mathbf{A}$, a block-column of $\mathbf{R}$, $\Lambda^{(k)}$, $[\Lambda^{(k)}]^{-1}$ are performed once per TDD burst, i.e. 100 times per second. The calculations $\mathbf{A}^H\mathbf{r}$, $F[\mathbf{A}^H\mathbf{r}]$, computing and inverse FFT of $[F(\hat{\mathbf{d}})]$ are performed twice per burst, i.e. 200 times per second. Four calculations are required to convert a complex operation into a real operation.

| Functions executed once per burst | MROPS |
|---|---|
| Calculating $\mathbf{A}$ | 3.0 |
| Calculating $\mathbf{A}^H\mathbf{A}$ | 4.4 |
| Calculating $F((\mathbf{R})_i)$ | 9.26 |
| Calculating $[\Lambda^{(k)}]^{-1}$ | 12.493 |

| Functions executed twice per burst | MROPS |
|---|---|
| Calculating $\mathbf{A}^H\mathbf{r}$ | 28.11 |
| Calculating $F[\mathbf{A}^H\mathbf{r}]$ | 2.3154 |
| Calculating $[F(\hat{\mathbf{d}})]_k = [\Lambda^{(k)}]^{-1}[F(\mathbf{A}^H\mathbf{r})]_k$ | 3.1232 |
| Calculating inverse FFT of $[F(\hat{\mathbf{d}})]$ | 2.3154 |

Thus, total number of MROPS for calculation efficient multiuser (joint) detection is 65.02 MROPS. Here $[\mathbf{A}^H\mathbf{r}]$ is calculated directly as a matrix-vector multiplication.

If FFTs are used to calculate $[\mathbf{A}^H\mathbf{r}]$ (using equation (36)), the computational complexity reduces from 65.02 to 63.99 MROPS. Though this reduction is not significant in the case of this novel fast joint detector, using equation (36) gives significant computational advantages over direct matrix-vector multiplication in the case of FFT based single-user detection algorithm SDFFT. Also, if a LU decomposition is used to determine $[\Lambda^{(k)}]^{-1}$, (using equations (38)-(40)), the calculation load of the novel FFT-based joint detector reduces to 54.87 MROPS.

A comparison of the complexity of fast joint detection and other detection techniques is now shown. The comparison of the following three data detection techniques for a TDD Burst Type I with SF = 16 and K = 8 is given below.

| Technique | MROPS |
| --- | --- |
| Approximate Cholesky based Joint Detector (JDChol) | 82.7 |
| Single User Detection: Approximate Cholesky based Equalization followed by a Hadamard transform based De-spreading (SDChol) | 205.23 |
| Fast Single User Detection: FFT based (SDFFT) | 69 |
| Fast Joint Detector (JDFFT) | 54.87 |

The complexity of SDChol is much higher than JDChol, because it involves chip level equalization. If for higher data rates, there are 12 codes of SF =16 each, the complexity of JDChol is 177 MROPS while the complexity of JDFFT is 90 MROPS, which is about 50% reduction in complexity.

**IV. D. Comparison with Single User Detection**

Performance of five algorithms will be compared for the downlink of a TD-CDMA system:

1) Approximate Cholesky based Joint Detector, denoted by *JDChol*, 2) Single User Detection: Approximate Cholesky based Equalization followed by Despreading, denoted by *SDChol,* 3) Single User Detection: FFT based Equalization followed by Despreading, denoted by *SDFFT* , 4) Matched Filter based Joint Detector denoted by *MF,* 5) FFT Based Joint Detector introduced in this paper, denoted by *JDFFT*.

JDChol is obtained by the application of approximate Cholesky based method in [3] to the joint detection problem in equation (6). The performance of the algorithms, in the multipath channels specified in TDD standards [1], is now analyzed. It will be seen in the simulations that single user detection algorithms like SDFFT, SDChol suffer from appreciable degradation in performance, compared to joint detector algorithms (JDChol, JDFFT) in some multipath channels. The degradation in performance of a FFT based single user detection algorithm (SDFFT) can be attributed to two factors

1. Degradation due to single user detection framework: In single user detection in the downlink, a mobile unit only uses its spreading code for decoding the data. This has an advantage in that there is no need to estimate the codes for the other users (referred to as blind code detection). However, this leads to a degradation in its performance, compared to joint detectors, which uses information about the spreading codes of all the users in its data estimation [2], [5].

2. Degradation due to using a FFT approximation:
The degradation in performance, due to the FFT approximation, can be quantified from equations (7) and (10). The number of block-rows of $\mathbf{R}$, that is outside the block-circulant structure, is equal to 2L, where L is the multipath delay spread (in symbols). When the maximum delay L of multipaths is small, the matrix $\mathbf{R}$ is close to being a circulant matrix. As the maximum delay L in the multipath profile increases (for a given power of the multipaths), the matrix $\mathbf{R}$ loses its circulant matrix, with 2LK of its rows deviating $\mathbf{R}$ from its circulant structure (this will be more relevant for Case 2 channel, as seen later).

The performance of the algorithms in different multipath channels is now analyzed. 3GPP Working Group 4 has specified multipath profiles of various fading channels. These multipath profiles are denoted by tddWg4Case1, tddWg4Case2 and tddWg4Case3. Case 1 and Case 3 channels have multipaths within 5 or 6 chip intervals. Case 3 channel is however a rapidly fading channel, the mobile speed being 120 km/hr. Case 2 channel however has equal power multipaths at delays of 1, 5 and 47 chips; the mobile speed is 3 km/hr.

First, the performance of the FFT based joint detector in this paper is compared to that of the approximate Cholesky based joint detector JDChol [2], [3]. Simulations indicate that for Case 1 and Case 3 channels, the JDFFT algorithm performs exactly the same as JDChol algorithm, at much reduced complexity. This is because in Case 1 and Case 3 channels, the multipath delays being small (i.e. W and L are small), the degradation due to the block-circulant approximation is very small (as explained above). Case 2 channel however has equal power multipaths with long delays, (i.e. W and L are large), which prevents the $\mathbf{R}$ from being block-circulant by a large margin. As mentioned earlier, the fraction of non-circulant rows in $\mathbf{R}$ is $(2L/N_s)$, which means that the fraction of non-circulant rows in R, for Case 2 channel (with a fixed number of symbols $N_s$ ), is more than that for Case 1 and Case 3 channels. Thus the degradation in performance for Case 2 channel (relative to JDChol) is appreciable (as seen from simulations). Simulations indicate that even for equal power multipaths, if the maximum delay spread L is reduced, the performance of JDFFT improves.

Single user detection algorithms also suffer from degradation due to factor number one. For this reason, performance comparisons are made with Cholesky based single user detector SDChol. The SDChol is shown only to illustrate the degradation due to factor number one only, as SDChol is not affected by the degradation due to the FFT approximation. It is seen that the performance of SDChol shows appreciable degradation, compared to JDChol as well as JDFFT. Comparisons are also made with FFT based single user detection algorithm SDFFT. This algorithm (though of reduced complexity) suffers from both the degradation due to single user detection framework, as well as due to the FFT approximation. Its performance is worse than JDChol, JDFFT and SDChol.

The development of JDFFT in this paper and [4] was motivated by the need to remove factor number one for degradation in performance, while achieving reduction in complexity as compared to JDChol.

**V. Simulation Results**

The performance of JDFFT, JDChol, SDChol, SDFFT along with matched filtering (MF), for the parameters of Burst Type I, is illustrated in the following plots. Channels specified by the WCDMA TDD specifications [1] were tested. For the approximate Cholesky based joint detector and the FFT based Joint Detector, it was assumed that the spreading codes of all the users are known. In general, the joint detector algorithms have to estimate the spreading codes of the other users, which adds to the complexity. All the simulations are for the downlink situation (without any transmit diversity). Simulation results are provided for the following cases. The simulations were performed over 800 timeslots.

In the first case, with spreading factor SF = 16 and number of users (codes) K = 8, Figures 2, 3 and 4 show the un-coded Bit Error Rate (BER) for Case 1, Case 3 and Case 2 channels respectively. For both Case 1 and Case 3 channels, the performance of JDFFT is very close to JDChol, which, for example, is a standard algorithm used in WCDMA TDD standard. Under the ``Single User Detection'' (SUD) category, even the Cholesky based SDChol, exhibits quite a bit of degradation compared to JDChol and JDFFT algorithms. FFT based single user detection SDFFT performs even worse.

Case 2 channel is characterized by equal power multipaths at delays of 1 and 5 chip intervals, and at a significant delay of 47 chip interval. A modification to the Case 2 channel has equal power multipaths at delays of 1, 5 and 9 (instead of 47) chip intervals. The simulation results are shown in Figure 5, where it is seen that JDFFT exhibits less degradation (with respect to JDChol) than in Figure 4, thereby illustrating that FFT based algorithms are susceptible to channels with multipaths at long delays.

The fast joint detector in this paper is suitable for higher sampling rates. Simulation results are provided, for oversampling by a factor of 2 and for high data rate services (2 Mbps data service) in Figures 6-8. Simulations also indicate that using a 64 point block FFT gives better performance than using a 61 point block FFT.

## VI. Conclusions

The single user detection method has an advantage in that it requires only the spreading code of the particular mobile unit at the downlink. It therefore obviates the need for blind code detection at the downlink of a TDD system. In this paper, a novel fast joint detector, based on block-FFT, is developed, which requires spreading codes of all the users. The novel JDFFT algorithm performs very close to JDChol, at much reduced complexity. Also JDFFT also shows a significant improvement in performance over single-user detection algorithms (SDChol and SDFFT), at lower complexity. FFT based joint detector (JDFFT) is thus a good choice in a TDD system and can be employed to deliver high data rate service in the downlink. The JDFFT algorithm can also be used for uplink applications. The novel method can also be applied to any generic multiple-input-multiple-output (MIMO) system like [18], [19].

$$\mathbf{R}_c\mathbf{D} =$$

$$\begin{bmatrix} (\mathbf{R}_0+\mathbf{R}_1+\mathbf{R}_2+\mathbf{R}_1^H+\mathbf{R}_2^H) & (\mathbf{R}_0 e^{j\frac{2\Pi}{N_s}}+\mathbf{R}_1^H e^{j\frac{4\Pi}{N_s}}+\mathbf{R}_2^H e^{j\frac{6\Pi}{N_s}}+\mathbf{R}_2 e^{j\frac{18\Pi}{N_s}}+\mathbf{R}_1 e^{j\frac{20\Pi}{N_s}}) & \cdots \\ \mathbf{R}_0+\mathbf{R}_1+\mathbf{R}_2+\mathbf{R}_1^H+\mathbf{R}_2^H) & (\mathbf{R}_1 e^{j\frac{2\Pi}{N_s}}+\mathbf{R}_0 e^{j\frac{4\Pi}{N_s}}+\mathbf{R}_1^H e^{j\frac{6\Pi}{N_s}}+\mathbf{R}_2^H e^{j\frac{8\Pi}{N_s}}+\mathbf{R}_2 e^{j\frac{20\Pi}{N_s}}) & \cdots \\ \vdots & \vdots & \vdots \\ \mathbf{R}_0+\mathbf{R}_1+\mathbf{R}_2+\mathbf{R}_1^H+\mathbf{R}_2^H) & \mathbf{R}_2 e^{j\frac{12\Pi}{N_s}}+\mathbf{R}_1 e^{j\frac{14\Pi}{N_s}}+\mathbf{R}_0 e^{j\frac{16\Pi}{N_s}}+\mathbf{R}_1^H e^{j\frac{18\Pi}{N_s}}+\mathbf{R}_2^H e^{j\frac{20\Pi}{N_s}}) & \cdots \\ \vdots & \vdots & \vdots \end{bmatrix}$$

$$\begin{bmatrix} (\mathbf{R}_0 e^{j\frac{2(N_s-1)\pi}{N_s}}+\mathbf{R}_1^H e^{j\frac{4(N_s-1)\pi}{N_s}}+\mathbf{R}_2^H e^{j\frac{6(N_s-1)\pi}{N_s}}+\mathbf{R}_2 e^{j\frac{2(N_s-1)^2\pi}{N_s}}+\mathbf{R}_1 e^{j\frac{2(N_s-1)N_s\pi}{N_s}}) \\ (\mathbf{R}_1 e^{j\frac{2(N_s-1)\pi}{N_s}}+\mathbf{R}_0 e^{j\frac{4(N_s-1)\pi}{N_s}}+\mathbf{R}_1^H e^{j\frac{6(N_s-1)\pi}{N_s}}+\mathbf{R}_2^H e^{j\frac{8(N_s-1)\pi}{N_s}}+\mathbf{R}_2 e^{j\frac{2(N_s-1)N_s\pi}{N_s}}) \\ \vdots \\ \mathbf{R}_2 e^{j\frac{2(N_s-1)(N_s-4)\pi}{N_s}}+\mathbf{R}_1 e^{j\frac{2(N_s-1)(N_s-3)\pi}{N_s}}+\mathbf{R}_0 e^{j\frac{2(N_s-1)(N_s-2)\pi}{N_s}}+\mathbf{R}_1^H e^{j\frac{2(N_s-1)^2\pi}{N_s}}+\mathbf{R}_2^H e^{j\frac{2(N_s-1)N_s\pi}{N_s}} \\ \vdots \end{bmatrix} \quad (14)$$

$$A_c D_2 = \begin{bmatrix} B(0) & 0 & 0 & 0 & \cdots & B(2) & B(1) \\ B(1) & B(0) & 0 & 0 & \cdots & & B(2) \\ B(2) & B(1) & B(0) & 0 & \cdots & & 0 \\ \vdots & \vdots & \vdots & \vdots & & \vdots & \vdots \\ 0 & 0 & \cdots & B(2) & B(1) & B(0) & \\ \vdots & \vdots & & & \vdots & & \end{bmatrix} \begin{bmatrix} I & e^{\frac{j2\pi}{N_s}} I_K & e^{\frac{j4\pi}{N_s}} I_K & \cdots & \cdots & e^{\frac{j2(N_s-1)\pi}{N_s}} I_K \\ I & e^{\frac{j4\pi}{N_s}} I_K & e^{\frac{j8\pi}{N_s}} I_K & \cdots & \cdots & e^{\frac{j4(N_s-1)\pi}{N_s}} I_K \\ \vdots & \vdots & \vdots & & & \vdots \\ I & e^{\frac{j2(N_s-1)\pi}{N_s}} I_K & e^{\frac{j4(N_s-1)\pi}{N_s}} I_K & \cdots & \cdots & e^{\frac{j2(N_s-1)^2\pi}{N_s}} I_K \\ I & e^{\frac{j2N_s\pi}{N_s}} I_K & e^{\frac{j4N_s\pi}{N_s}} I_K & \cdots & \cdots & e^{\frac{j2(N_s-1)N_s\pi}{N_s}} I_K \end{bmatrix} =$$

$$\begin{bmatrix} (B(0)+B(1)+B(2)) & (B(0)e^{\frac{j2\pi}{N_s}} + B(1)e^{\frac{j2N_s\pi}{N_s}} + B(2)e^{\frac{j2(N_s-1)\pi}{N_s}}) & \cdots & (B(0)e^{\frac{j2(N_s-1)\pi}{N_s}} + B(1)e^{\frac{j2N_s(N_s-1)\pi}{N_s}} + B(2)e^{\frac{j2(N_s-1)^2\pi}{N_s}}) \\ (B(0)+B(1)+B(2)) & (B(0)e^{\frac{j4\pi}{N_s}} + B(1)e^{\frac{j2\pi}{N_s}} + B(2)e^{\frac{j2N_s\pi}{N_s}}) & \cdots & (B(0)e^{\frac{j4(N_s-1)\pi}{N_s}} + B(1)e^{\frac{j2(N_s-1)\pi}{N_s}} + B(2)e^{\frac{j2(N_s-1)N_s\pi}{N_s}}) \\ \vdots & \vdots & \cdots & \vdots \\ (B(0)+B(1)+B(2)) & (B(0)e^{\frac{j2N_s\pi}{N_s}} + B(1)e^{\frac{j2(N_s-1)\pi}{N_s}} + B(2)e^{\frac{j2(N_s-2)\pi}{N_s}}) & \cdots & (B(0)e^{\frac{j2(N_s-1)N_s\pi}{N_s}} + B(1)e^{\frac{j2(N_s-1)^2\pi}{N_s}} + B(2)e^{\frac{j2(N_s-1)(N_s-2)\pi}{N_s}}) \\ \vdots & \vdots & & \vdots \end{bmatrix} .(32)$$

**Figure 1**     **COMMUNICATION BURST**

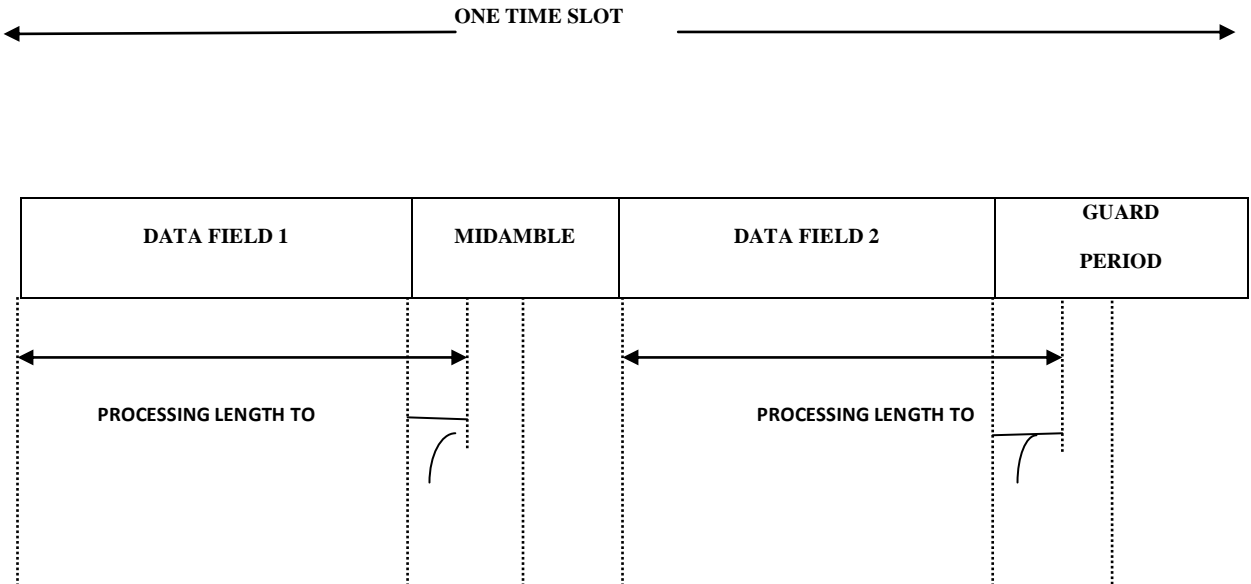

INCLUDE LAST SYMBOLS 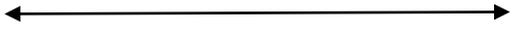 INCLUDE LAST SYMBOLS 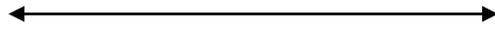

IMPULSE RESPONSE         W-1                    IMPULSE RESONSE          W-1

PROCESSING LENGTH TO HAVE            PROCESSING LENGTH TO HAVE

DESIRED PFA LENGTH                   DESIRED PFA LENGTH

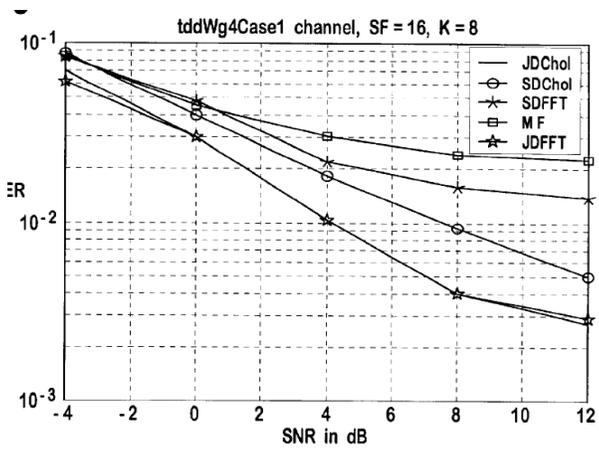

**Figure 2** Case 1 channel, SF =16, K=8.

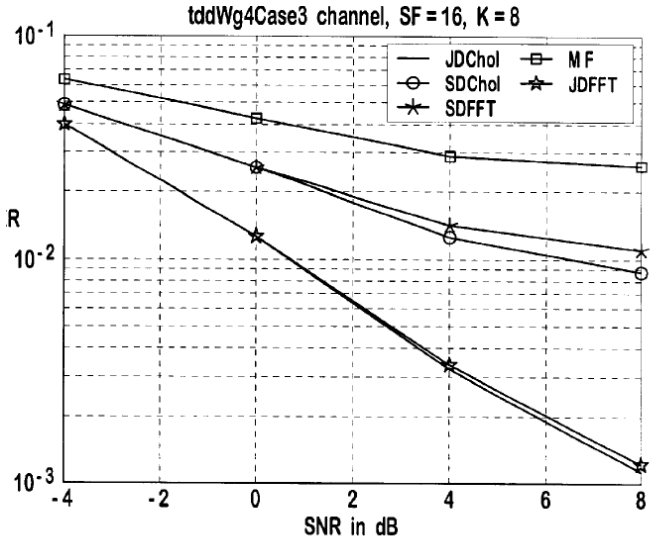

**Figure 3** Case 3 channel, SF =16, K=8.

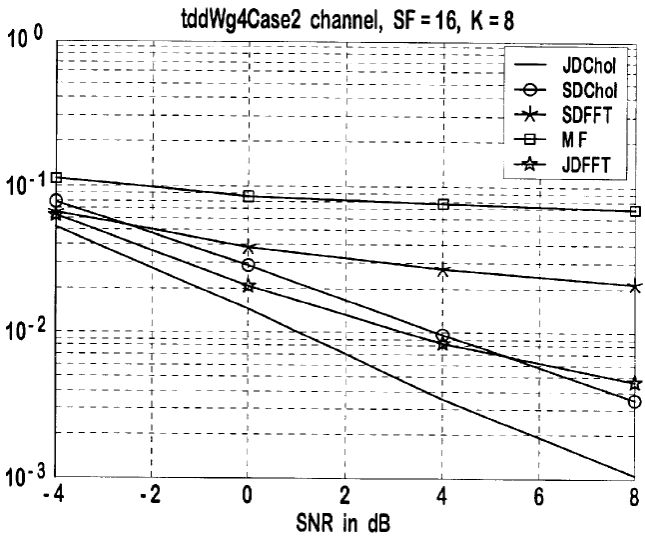

**Figure 4** Case 2 channel, SF =16, K=8.

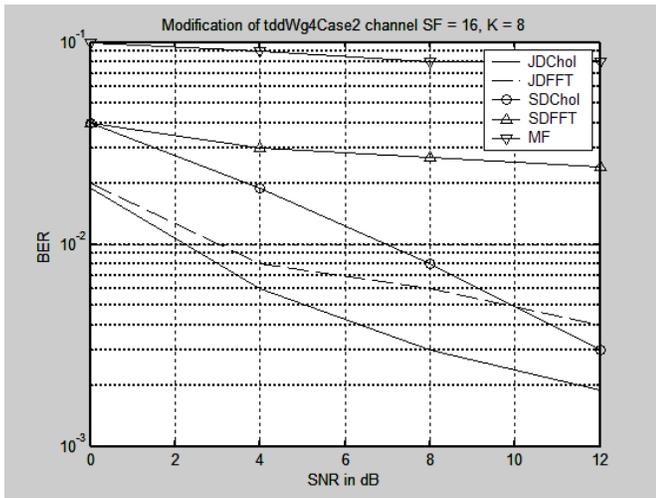

**Figure 5**  Modification of Case 2 channel, SF =16, K=8.

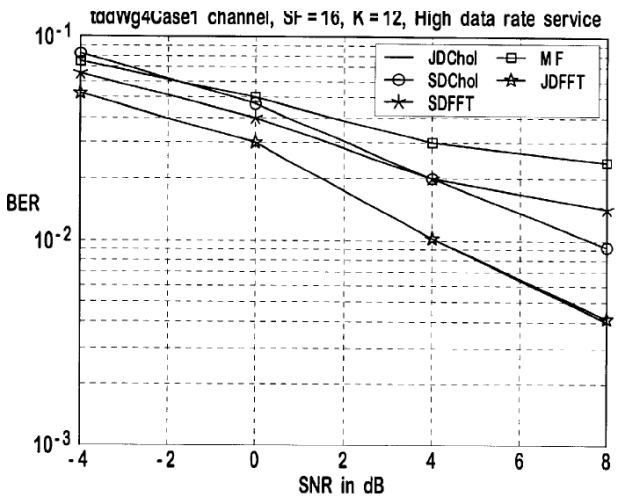

**Figure 6**  Oversampled case, Case 1 channel. High data rate service.

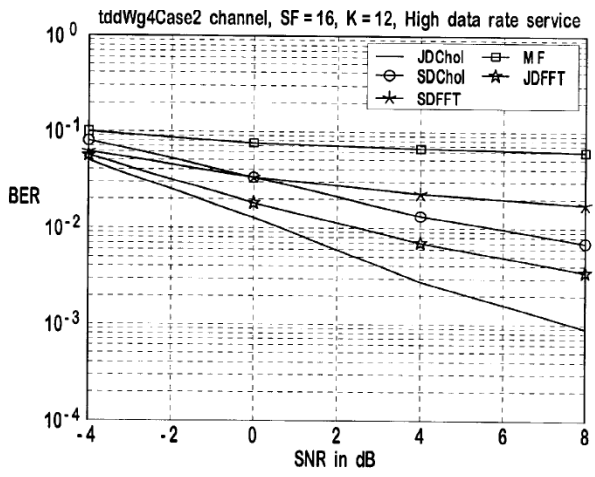

**Figure 7**   Case 2 channel, SF =16, 12 codes, High data rate service.  Oversampled case.

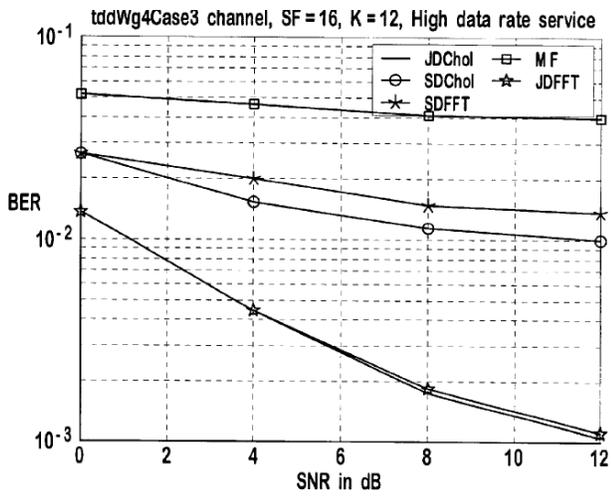

**Figure 8** Case 3 channel, SF =16, 12 codes, High data rate service.   Oversampled case.